\documentclass[aps,showpacs,twocolumn]{revtex4}
\usepackage{epsfig}

\begin{document}

\title{QCD quark cyclobutadiene and light tetraquark spectra
 \footnote{Email address: jlping@njnu.edu.cn (J.L. Ping)}}

\author{Chengrong Deng$^{a}$, Jialun Ping$^b$, Hui Wang$^b$, Ping Zhou$^{a}$, Fan Wang$^c$}

\affiliation{$^a$School of Mathematics and Physics, Chongqing
Jiaotong University, Chongqing 400074, P.R. China}

\affiliation{$^b$Department of Physics, Nanjing Normal University,
Nanjing 210097, P.R. China}

\affiliation{$^c$Department of Physics, Nanjing University,
Nanjing 210093, P.R. China}

\begin{abstract}
The QCD quark cyclobutadiene (ring-like), a new color structure of
tetraquark system, is proposed and studied in the flux-tube model
with a multi-body confinement interaction. Numerical calculations
show that the light tetraquark systems ($u, d, s$ only) with
cyclobutadiene, diquark-antidiquark flux-tube structures have
similar energies and they can be regarded as QCD isomeric
compounds. The energies of some tetraquark states are close to the
masses of some excited mesons and so in the study of these mesons,
the tetraquark components should be taken into account. There are
also some meson states, $\sigma$, $\kappa(800)$, $f_0(980)$,
$f_0(1500)$, $\pi_1(1400)$, $\pi_1(1600)$, $f_2(1430)$ and
$K^*(1410)$, where tetraquark components might be dominant. The
meson states with exotic quantum numbers are studied as the
tetraquark states. The multi-body confinement interaction reduces
the energy of the tetraquark state in comparison with the usual
additive two body confinement interaction model.

\end{abstract}


\maketitle

\section{Introduction}

In the constituent quark model (CQM), mesons are assumed to be
composed of $q\bar{q}$. Although various properties of light mesons
have been explained within this $q\bar{q}$ minimum Fock space, there
are still properties of some meson states can not be described well
by this quark model~\cite{godfrey,amsler,jaffe}. In fact, mesons
might be more complicated objects with higher Fock components other
than the lowest $q\bar{q}$. The wavefunction of zero baryon number
(B=0) hadron, if the gluon degree of freedom is neglected, should be
in general as
\begin{equation}
|B=0\rangle=\sum_{n=1}c_n|q^n\bar{q}^n\rangle,
\end{equation}
where $n=1,2,3,\cdots$. The recent studies on meson spectroscopy
called for unquench the quark model, i.e., the $q\bar{q}$ and
$q^2\bar{q}^2$ mixing~\cite{jaffe,aubert,choi,evdokimov}.
Furthermore the introduction of tetraquark states $q^2\bar{q}^2$
is indispensable for the states with exotic quantum
numbers~\cite{thompson,aele,adams,alekseev,nerling}. In recent
years, comprehensive researches on tetraquark states have been
carried out by many
authors~\cite{wong,close,swanson,tornqvist,maiani1,hogaasen,ebert1,barnea,vijande1,janc},
because Belle, BaBar and other experimental collaborations have
observed many open and hidden charmed hadrons, which are difficult
to be fitted into the conventional meson $c\bar{c}$
spectra~\cite{olson}. The states with quantum numbers
$J^{PC}=0^{--}$, $\mbox{even}^{+-}$ and $\mbox{odd}^{-+}$ had been
theoretically studied as tetraquark
states~\cite{hxchen,ckjiao,dudek}. Experimental evidences of the
exotic states with quantum numbers $J^{PC}=1^{-+}$ were
accumulated~\cite{thompson,aele,adams,alekseev,nerling}. The
investigations of multiquark states with flux-tube structures will
provide important low energy quantum chromodynamics (QCD)
information, such as $q\bar{q}\bar{q}$ and $qq\bar{q}$
interactions~\cite{dmitrasinovic}, which is absent in ordinary
hadrons due to their unique flux-tube structure.

QCD is widely accepted as the fundamental theory of the strong
interaction, in which color confinement is a long-distance
behavior whose understanding continues to be a challenge for
theoretical physics. Lattice QCD (LQCD) allows us to investigate
the confinement phenomenon in a nonperturbative framework and its
calculations on mesons, baryons, tetraquark and pentaquark states
reveal flux-tube or string like
structure~\cite{alexandrou,takahashi,okiharu1,okiharu2}. Such
flux-tube like structures lead to a ``phenomenological''
understanding of color confinement and naturally to a linear
confinement potential in $q\bar{q}$ and $q^3$ quark systems.

It is well known that nuclear force and molecule force are very
similar except for the length and energy scale
difference~\cite{anderson,teng}. For multi-body systems, the
flux-tubes in a multi-quark system should be also very similar to
the chemical bond in the molecular system. Among organic
compounds, the same molecular constituents may have different
chemical bond structure, which are named as isomeric compounds. In
hadronic world, the multiquark states with same quark contents but
different flux-tube structures should be similarly called as QCD
isomeric compounds. The past theoretical studies on multiquark
states reveal various flux-tube
structures~\cite{barnes,lee,xliu,jaffe1,maiani2,ebert2,hxhuang,jlping,ping,ppbar}:
hadron molecular states $[q\bar{q}]_1[q\bar{q}]_1$,
$[q\bar{q}]_1[q^3]_1$, $[q^3]_1[q^3]_1$ and $[\bar{q}^3]_1[q^3]_1$
and hidden color states $[[q\bar{q}]_8[q\bar{q}]_8]_1$,
$[[q^2]_{\bar{3}}[\bar{q}^2]_3]_1$, $[[q\bar{q}]_8[q^3]_8]_1$,
$[[q^2]_{\bar{3}}[q^2]_{\bar{3}}\bar{q}]_1$, $[[q^4]_3\bar{q}]_1$,
$[[q^3]_8[q^3]_8]_1$,
$[[q^2]_{\bar{3}}[q^2]_{\bar{3}}[q^2]_{\bar{3}}]_1$,
$[[\bar{q}^3]_8[q^3]_8]_1$ and a QCD quark benzene $[q^6]_1$, et
al, here the subscripts represent color dimensions, which should
be mixed and affect the corresponding hadron properties if they
really exist.

Based on the chemical benzene and the similarity between color
flux-tubes and chemical bonds, a new flux-tube structure, the
quark benzene, for a six-quark system was proposed and its
possible effect on $NN$ scattering was discussed in our previous
paper~\cite{ping}. In the present work, a new flux-tube structure
for a tetraquark state, which is similar to the molecular
cyclobutadiene and is therefore called as a QCD quark
cyclobutadiene, is proposed. The aims of this paper are: (i) to
investigate the properties of a QCD cyclobutadiene in the
flux-tube model, this model involves a multibody confinement
potential and has been successfully applied to multiquark
systems~\cite{deng,ppbar}; (ii) to study the spectra of light
tetraquark states with two flux-tube structures
(diquark-antidiquark and QCD quark cyclobutadiene), which helps us
to understand the meson states beyond a $q\bar{q}$ configuration
and will provide a new sample to study the mixing of $q\bar{q}$
and $qq\bar{q}\bar{q}$. The paper is organized as follows: four
possible flux-tube structures of a tetraquark system are discussed
in Sec. II. Section III is devoted to the description of the
flux-tube model and the multibody confinement potentials of a
diquark-antidiquark and a QCD quark cyclobutadiene structures. A
brief introduction of the construction of the wave functions and
quantum numbers of a tetraquark state are given in Sec. IV. The
numerical results and discussions are presented in Section V. A
brief summary is given in the last section.

\section{Flux-tube structures of a tetraquark state}

In the flux-tube picture it is assumed that the color-electric
flux is confined to narrow, flux-tube like tubes joining quarks. A
flux-tube starts from every quark and ends at an antiquark or a
Y-shaped junction, where three flux-tubes annihilate or be
created~\cite{isgur}. In general, a state with $N+1$-particles can
be generated by replacing a quark or an antiquark in an
$N$-particles state by a Y-shaped junction and two antiquarks or two
quarks. According to this point of view, there are four
possible flux-tube structures for a tetraquark system as shown in
Fig.1, where $\mathbf{r}_i$ represents the position of a quark
$q_i$ (antiquark $\bar{q}_i$) which is denoted by a solid (hollow)
dot, $\mathbf{y}_i$ represents a junction where three flux-tubes
meet. A thin line connecting a quark and a junction represents a
fundamental flux tube, {\em i.e.} color triplet. A thick line
connecting two junctions is for a color sextet, octet or others,
namely a compound flux tube. The numbers on the flux-tubes
represent the color dimensions of the corresponding flux tube. The
different types of flux-tube may have different
stiffness~\cite{Bali}, the detail will be discussed in the next
section. Both the overall color singlet nature of a multi-quark
system and the $SU(3)$ color coupling rule at each junction must
be satisfied.

The flux-tube structure (\textbf{a}) in Fig. 1 is a meson-meson
molecule state, many newly observed exotic hadrons are discussed
in this picture~\cite{barnes,lee,xliu}. The tetraquark states with
the flux-tubes tructure (\textbf{b}) generally have high energies
due to a repulsive interaction between a quark and an antiquark in
a color octet meson. Thus this flux-tube structure is often
neglected in the study of multiquark states. However sometimes the
attraction between two color octet mesons will lower the energies
of the system considerably. In the case of the flux-tube structure
(\textbf{c}), called as a diquark-antidiquark structure, it has
two possible color coupling schemes, namely
$[[qq]_{\bar{3}}[\bar{q}\bar{q}]_{3}]_1$ and
$[[qq]_6[\bar{q}\bar{q}]_{\bar{6}}]_1$, the latter is expected to
be a highly excited state and therefore the $6$ diquark is usually
named as a ``bad'' diquark, since the interaction between two
symmetric quarks (antiquarks) is repulsive, thus many authors are
in favor of the $\bar{3}$ (``good") diquark
picture~\cite{jaffe1,maiani2,ebert2}.

\begin{figure}
 \epsfxsize=3.8in \epsfbox{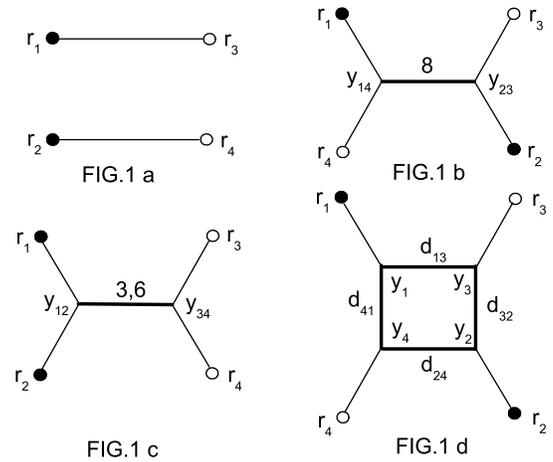}
\caption{Four possible flux-tube structures.}
\end{figure}

The first three flux-tube structures can be explained as the basic
structures for a tetraquark system. The last structure can be
generated by means of exciting two Y-shape junctions and three
compound flux-tubes from vacuum based on the second or third
structures. In the constituent quark model, a quark is massive.
One can suppose that the recombination of flux-tubes is faster
than the motion of the quarks. Subsequently, the ends of four
compound flux-tubes meet each other in turn to form a closed
flux-tube structure, a
ring-$\mathbf{y}_1\mathbf{y}_3\mathbf{y}_2\mathbf{y}_4$, which was
interpreted as a pure gluon state by Isgur and Paton~\cite{isgur},
as a glueball state, and discussed in the framework of the dual
Ginzburg-Landau theory~\cite{koma}. With quarks or antiquarks
connecting to vertexes of the square by a fundamental flux tube,
this picture could be explained as a $qq\bar{q}\bar{q}$-glueball
hybrid. According to overall color singlet and $SU(3)$ color
coupling rule, the corresponding compound color flux-tube
dimensions $(d_{13}, d_{32}, d_{24}, d_{41})$ have six different
sets: $(3, 8, 3, 8)$, $(\bar{6}, 8, \bar{6}, 8)$, $(\bar{3}, 3,
\bar{3}, 3)$, $(8, \bar{3}, 8, \bar{3})$, $(8, 6, 8, 6)$ and
$(\bar{3}, \bar{6},\bar{3},\bar{6})$. The flux-tubes locate in
opposite sides of the
ring-$\mathbf{y}_1\mathbf{y}_3\mathbf{y}_2\mathbf{y}_4$ have the
same dimension, which is similar to the symmetry with the
distribution of double bonds and single bonds in a cyclobutadiene
in chemistry. We thus name the flux-tube structure (\textbf{d}) as
a QCD quark cyclobutadiene. Of course, the existence of another
QCD quark cyclobutadiene is also allowed in which two quarks or
antiquarks seat neighboring positions in the flux-tubering.
Certainly, more complicated configuration are permitted, including
more Y-shaped junctions and more complex topological structures.

\section{the flux-tube model and multi-body confinement potentials}

Recently, LQCD and nonperturbative QCD method have made impressive
progresses on hadron properties, even on hadron-hadron
interactions~\cite{maris,ishii,takahashi1,inoue,ishii1}. However,
QCD-inspired CQM is still an useful tool in obtaining physical
insight for these complicated strong interaction systems. CQM can
offer the most complete description of hadron properties and is
probably the most successful phenomenological model of hadron
structure~\cite{godfrey}. In traditional CQM, a two-body
interaction proportional to the color charges
$\mathbf{\lambda}_i\cdot\mathbf{\lambda}_j$ and $r_{ij}^n$, where
$n=1$ or 2 and $r_{ij}$ is the distance between two quarks, was
introduced to phenomenologically describe quark confinement
interaction. The traditional model can well describe the
properties of ordinary hadrons ($qqq$ and $q\bar{q}$) because the
flux-tube structures for an ordinary hadron are unique and
trivial. However, the traditional model is known to be flawed
phenomenologically because it leads to power law van der Waals
forces between color-singlet
hadrons~\cite{feinberg,greenberg,weinstein1}. It is also flawed
theoretically in that it is very implausible that the long-range
static multi-body potential is just a sum of the two-body
ones~\cite{weinstein}. Many papers were devoted to eliminate the
physically nonexisting long-distance van der Waals force arising
from the traditional models based on the sum of two-body Casimir
scaled potentials~\cite{oka1,oka2,karliner,vijande2}.

LQCD studies show that the confinement potential of a multiquark
state is a multibody interaction which is proportional to the
minimum of the total length of flux-tubes which connects the
quarks to form a multiquark
state~\cite{alexandrou,takahashi,okiharu1,okiharu2}. The naive
flux-tube model is developed based on LQCD picture by taking into
account a multi-body confinement potential with a harmonic
interaction approximation, i.e., a sum of the square of the length
of flux-tubes rather than a linear one is assumed to simplify the
calculation~\cite{ping,ww}. The approximation is justified with
the following two reasons: one is that the spatial variations in
separation of the quarks (lengths of the flux tube) in different
hadrons do not differ significantly, so the difference between the
two functional forms is small and can be absorbed in the
adjustable parameter, the stiffness. The other is that we are
using a nonrelativistic dynamics in the study. As was shown long
ago~\cite{goldman}, an interaction energy that varies linearly
with separation between fermions in a relativistic first order
differential dynamics has a wide region in which a harmonic
approximation is valid for the second order (Feynman-Gell-Mann)
reduction of the equations of motion. Combining with Gaussian
expansion method (GEM), the flux-tube model including one gluon
exchange and one boson exchange interactions was successfully
applied to new hadronic states and some interesting results
were obtained~\cite{deng,ppbar}.

Within the flux-tube picture, the flux-tubes in the
ring-structure, see Fig.1d, are assumed to have the same
properties as the flux-tubes in the ordinary meson or
baryon~\cite{Nawa}. Thus in the flux-tube model with quadratic
confinement, the confinement potentials $V^c$ and $V^d$ for
diquark-antidiquark and cyclobutadiene structures,
have the following forms, respectively.
\begin{eqnarray}
V^c &=&k\left[ (\mathbf{r}_1-\mathbf{y}_1)^2
+(\mathbf{r}_2-\mathbf{y}_1)^2+(\mathbf{r}_3-\mathbf{y}_2)^2\right. \nonumber \\
&+&
\left.(\mathbf{r}_4-\mathbf{y}_2)^2+\kappa_{d_{12}}(\mathbf{y}_1-\mathbf{y}_2)^2\right], \\
V^d &=& k\left[ \sum_{i=1}^4(\mathbf{r}_{i}-\mathbf{y}_{i})^2
 +{\sum_{i<j}}^{\prime}
  \kappa_{d_{ij}}(\mathbf{y}_i-\mathbf{y}_j)^2 \right]
\end{eqnarray}
where the $\sum^{\prime}$ means that the summation is over the
adjacent junction pairs on a compound flux tube, this term is the
energy of the flux-tube
ring-$\mathbf{y}_1\mathbf{y}_3\mathbf{y}_2\mathbf{y}_4$. The
parameter $k$ is the stiffness of an elementary flux tube, while
$k\kappa_{d_{ij}}$ is other compound flux-tube stiffness. The
compound flux-tube stiffness parameter $\kappa_{d_{ij}}$ depends
on the color dimension, $d_{ij}$, of the flux tube~\cite{Bali},
\begin{equation}
 \kappa_{d_{ij}}=\frac{C_{d_{ij}}}{C_3},
\end{equation}
where $C_{d_{ij}}$ is the eigenvalue of the Casimir operator
associated with the $SU(3)$ color representation $d_{ij}$ on
either end of the flux tube, namely $C_3=\frac{4}{3}$,
$C_6=\frac{10}{3}$ and $C_8=3$.

For given quark positions $\mathbf{r}_i$, the positions of those
junctions $\mathbf{y}_i$, variational parameters, can be
determined by means of minimizing the confinement potentials $V^c$
and $V^d$. To simplify the formats of $V^c$ and $V^d$ after
obtaining the positions of the junctions $\mathbf{y}_i$, two set
of canonical coordinates $\mathbf{R}_i$ and $\mathcal{R}_i$ can be
introduced respectively and written as,
\begin{eqnarray}
\left (
\begin{array}{ll}
\mbox{$\mathbf{R}_1$}\\
\mbox{$\mathbf{R}_2$}\\
\mbox{$\mathbf{R}_3$}\\
\mbox{$\mathbf{R}_4$}\\
\end{array}
\right )
 =\left (
\begin{array}{ll}
\mbox{$\frac{1}{\sqrt{2}}$~~$\frac{-1}{\sqrt{2}}$~~~~0~~~~0}\\
\mbox{~0~~~~0~~~~$\frac{1}{\sqrt{2}}$~~$\frac{-1}{\sqrt{2}}$}\\
\mbox{$\frac{1}{\sqrt{4}}$~~$\frac{1}{\sqrt{4}}$~~$\frac{-1}{\sqrt{4}}$~~$\frac{-1}{\sqrt{4}}$}\\
\mbox{$\frac{1}{\sqrt{4}}$~~$\frac{1}{\sqrt{4}}$~~$\frac{1}{\sqrt{4}}$~~$\frac{1}{\sqrt{4}}$}\\
\end{array}
\right ) \left (
\begin{array}{ll}
\mbox{$\mathbf{r}_1$}\\
\mbox{$\mathbf{r}_2$}\\
\mbox{$\mathbf{r}_3$}\\
\mbox{$\mathbf{r}_4$}\\
\end{array}
\right )\\
\left (
\begin{array}{ll}
\mbox{$\mathcal{R}_1$}\\
\mbox{$\mathcal{R}_2$}\\
\mbox{$\mathcal{R}_3$}\\
\mbox{$\mathcal{R}_4$}\\
\end{array}
\right )
 =\left (
\begin{array}{ll}
\mbox{$\frac{1}{\sqrt{4}}$~~$\frac{-1}{\sqrt{4}}$~~$\frac{-1}{\sqrt{4}}$~~$\frac{1}{\sqrt{4}}$}\\
\mbox{$\frac{1}{\sqrt{4}}$~~$\frac{1}{\sqrt{4}}$~~$\frac{-1}{\sqrt{4}}$~~$\frac{-1}{\sqrt{4}}$}\\
\mbox{$\frac{1}{\sqrt{4}}$~~$\frac{-1}{\sqrt{4}}$~~$\frac{1}{\sqrt{4}}$~~$\frac{-1}{\sqrt{4}}$}\\
\mbox{$\frac{1}{\sqrt{4}}$~~$\frac{1}{\sqrt{4}}$~~$\frac{1}{\sqrt{4}}$~~$\frac{1}{\sqrt{4}}$}\\
\end{array}
\right ) \left (
\begin{array}{ll}
\mbox{$\mathbf{r}_1$}\\
\mbox{$\mathbf{r}_2$}\\
\mbox{$\mathbf{r}_3$}\\
\mbox{$\mathbf{r}_4$}\\
\end{array}
\right )
\end{eqnarray}
The minimums $V_{min}^c$ and $V_{min}^d$ of the confinement
potentials can be divided into three independence harmonic
oscillators and have therefore the following forms,
\begin{eqnarray}
V^{c}_{min} &=& k\left[\mathbf{R}_1^2+\mathbf{R}_2^2+
\frac{\kappa_{d_{12}}}{1+\kappa_{d_{12}}}\mathbf{R}_3^2\right]\\
V_{min}^d & = & k\left[\frac{2\kappa_{d_1}}{1+2\kappa_{d_1}}
\mathcal{R}_{1}^2 +\frac{2\kappa_{d_2}}{1+2\kappa_{d_2}}
\mathcal{R}_{2}^2  \right. \nonumber \\
&+& \left. \frac{2(\kappa_{d_1}+\kappa_{d_2})}
{1+2(\kappa_{d_1}+\kappa_{d_2})}\mathcal{R}_{3}^2 \right],
\end{eqnarray}
where the parameters $\kappa_{d_1}$ and $\kappa_{d_2}$ are used to
describe the stiffness of two sets of opposite flux-tubes in the
ring-$\mathbf{y}_1\mathbf{y}_3\mathbf{y}_2\mathbf{y}_4$ due to the
symmetry, respectively. Obviously, the confinement potentials
$V^c_{min}$ and $V^d_{min}$ are multi-body interactions rather
than the sum of two-body interactions.

The limit $\kappa_{d_{ij}}$ going to infinity indicates that the
corresponding compound flux-tube contracts to a junction due to
the requirement of the minimum of the confinement. The limit
$\kappa_{d_{ij}}$ going to zero indicates the rupture of the
corresponding compound flux-tube and then a multiquark state
decays into several color singlet hadrons. The flux-tube
structures of a multiquark state can therefore change if the
$\kappa_{d_{ij}}$ is taken as an adjustable parameter. In the
limit $\kappa_{d_1}$ or $\kappa_{d_2}$ going to infinity, a QCD
quark cyclobutadiene reduces to a two-color-octet meson state or a
diquark-antidiquark state. In the limit one of $\kappa_1$ and
$\kappa_2$ going to infinity and the other going to zero, a QCD
quark cyclobutadiene decays into two color mesons. In the limit
all $\kappa_d$'s in Fig.1 going to infinity, the last three
flux-tube structures reduce to one structure due to all compound
flux-tubes shrink to a junction, leaving a hub and spokes
configuration.

Taking into account a potential energy shift $\Delta$ in each
independent harmonic oscillator, the confinement potentials
$V_{min}^c$ and $V_{min}^d$ have therefore the following forms
\begin{eqnarray}
V^c_{min} &=&
k\left[(\mathbf{R}_1^2-\Delta)+(\mathbf{R}_2^2-\Delta)+
\frac{\kappa_{d_{12}}}{1+\kappa_{d_{12}}}(\mathbf{R}_3^2-\Delta)\right]\nonumber\\
\\ V_{min}^d & = & k\left[\frac{2\kappa_{d_1}}{1+2\kappa_{d_1}}
(\mathcal{R}_{1}^2-\Delta) +\frac{2\kappa_{d_2}}{1+2\kappa_{d_2}}
 (\mathcal{R}_{2}^2-\Delta)  \right. \nonumber \\
&+& \left. \frac{2(\kappa_{d_1}+\kappa_{d_2})}
{1+2(\kappa_{d_1}+\kappa_{d_2})} (\mathcal{R}_{3}^2-\Delta)
\right],
\end{eqnarray}
where the parameters $k$ and $\Delta$ are determined by fitting
ordinary meson spectra~\cite{deng}. Carlson and Pandharipande also
considered similar flux-tube energy shift which is proportional to
the number of quarks $N$~\cite{carlson}.

One gluon exchange and one Goldstone boson exchange interactions
not only are important and responsible for the mass splitting in
the meson spectra but also are indispensable for the
investigations on the multiquark system~\cite{ppbar}, the details
of the parts of model Hamiltonian can be found in our previous
paper~\cite{deng}.

\section{wave functions and definition of quantum numbers}

The flux-tube structure specifies how the colors of quarks and
anti-quarks are coupled to form an overall color singlet. It is
however difficult to construct the color wave function of the QCD
quark cyclobutadiene only using quark degrees of freedom in the
framework of the quark models. In order to comprehensively study a
QCD quark cyclobutadiene, one gluon exchange and one boson
exchange interactions have to be included. The color wave function
of a QCD quark cyclobutadiene is therefore indispensable and
approximately assumed to be the same as that of a
diquark-antidiquark structure. In the framework of a
diquark-antidiquark structure, three relative motions are shown in
FIG. 2, where $\mathbf{r}_i$ represents the position of the quark
$q_i$ (antiquark $\bar{q}_i$) which is denoted by a solid (hollow)
dot, the corresponding Jacobi coordinates can be expressed as
\begin{eqnarray}
\mathbf{r}&=&\mathbf{r}_1-\mathbf{r}_2,\ {}
\mathbf{R}=\mathbf{r}_3-\mathbf{r}_4,\nonumber\\
\mathbf{X}&=&\frac{m_1\mathbf{r}_1+m_2\mathbf{r}_2}{m_1+m_2}
-\frac{m_3\mathbf{r}_3+m_4\mathbf{r}_4}{m_3+m_4},
\end{eqnarray}
$L$, $l_1$ and $l_2$ are the orbital angular momenta associated
with the relative motion coordinates $\mathbf{X}$, $\mathbf{r}$
and $\mathbf{R}$, respectively. The total wave function of a
tetraquark state can be written as a sum of the following direct
products of color, isospin, spin and spatial terms,
\begin{eqnarray}
\Phi^{q^2\bar{q}^2}_{IJ}\!\! &=& \!\!
\sum_{\alpha}\xi_{\alpha}^{IJ}\left[ \left[
\left[\phi^G_{l_1}(\mathbf{r})\chi_{s_1}\right]_{J_1}
\left[\psi^G_{l_2}(\mathbf{R})\chi_{s_2}\right]_{J_2}
\right ]_{J_{12}} \right. \nonumber \\
& \times & \left. F^G_{L}(\mathbf{X})\right]_{J}
\left[\eta_{I_1}\eta_{I_2}\right]_{I}
\left[\chi_{c_1}\chi_{c_2}\right]_{C},
\end{eqnarray}
where $I$ and $J$ are total isospin and angular momentum. $\alpha$
represents all possible intermediate quantum numbers
$\alpha=\{l_i,s_i,J_i,J_{12},L,I_i\}$, where $i=1,2$.
$\chi_{s_i}$, $\eta_{I_i}$ and $\chi_{c_i}$ are spin, flavor and
color wave functions of diquark or anti-diquark, respectively.
[~]'s denote Clebsh-Gordan coefficients coupling. The overall
color singlet can be constructed in two ways,
$\chi_c^1=\bar{3}_{12}\otimes3_{34}$,
$\chi_c^2=6_{12}\otimes\bar{6}_{34}$, ``good" diquark and ``bad"
diquark are both included. Taking into account all degrees of
freedom, the Pauli principle must be satisfied for each subsystem
of identical quarks or antiquarks. The coefficient
$\xi^{IJ}_{\alpha}$ is determined by diagonalizing the
Hamiltonian.
\begin{figure}
 \epsfxsize=1.8in \epsfbox{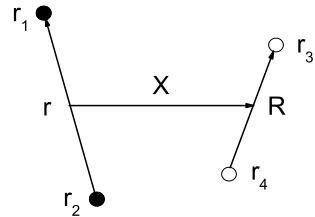}
\caption{Jacobi coordinates for a $q^2\bar{q}^2$ system}
\end{figure}

To obtain a reliable solution of a few-body problem, a high
precision method is indispensable. In this work, a Gaussian
Expansion Method (GEM)~\cite{GEM}, which has been proven to be
rather powerful in solving a few-body problem, is used to study
four-body systems in the flux-tube model. In GEM, three relative
motion wave functions are expanded as,
\begin{eqnarray}
\phi^G_{l_1m_1}(\mathbf{r}) & =& \sum_{n_1=1}^{n_{1max}}
 c_{n_1}N_{n_1l_1}r^{l_1}
 e^{-\nu_{n_1}r^2}Y_{l_1m_1}(\hat{\mathbf{r}}),
  \nonumber \\
\psi^G_{l_2m_2}(\mathbf{R}) & =& \sum_{n_2=1}^{n_{2max}}
 c_{n_2}N_{n_2l_2}R^{l_2}
 e^{-\nu_{n_2}R^2}Y_{l_2m_2}(\hat{\mathbf{R}}),
 \\
F^G_{LM}(\mathbf{X}) & =& \sum_{n_3=1}^{n_{3max}}
c_{n_3}N_{LM}X^{L}e^{-\nu_{n_3}X^2}Y_{LM}(\hat{\mathbf{X}}),\nonumber
\end{eqnarray}
where $N_{n_1l_1}$, $N_{n_2l_2}$ and $N_{n_3l_3}$ are
normalization constants. Gaussian size parameters are taken as the
following geometric progression numbers
\begin{eqnarray}
\nu_{n}=\frac{1}{r^2_n},& r_n=r_1a^{n-1},&
a=\left(\frac{r_{n_{max}}}{r_1}\right)^{\frac{1}{n_{max}-1}}.
\end{eqnarray}

The parity for a diquark-antidiquark state is the product of the
intrinsic parities of two quarks and two antiquarks times the
factors coming from the spherical harmonics~\cite{Llanes-Estrada}
\begin{eqnarray}
P=P_qP_qP_{\bar{q}}P_{\bar{q}}(-1)^{l_1+l_2+L}=(-1)^{l_1+l_2+L}.
\end{eqnarray}
Using our coordinates, the eigenvalues of the charge conjugation
of a diquark-antidiquark state can be calculated by following the
same steps as in the $q\bar{q}$ case. We can consider a
diquark-antidiquark state as a $Q\bar{Q}$ meson, where $\bar{Q}$
and $Q$ represent a diquark and an antidiquark, respectively, with
total ``spin'' $J_{12}$ and relative angular momentum $L$ between
$Q$ and $\bar{Q}$ (see Eq. 8). The $C$-parity eigenvectors are
those states for which $Q$ and $\bar{Q}$ have opposite charges. So
applying the charge conjugation operator to these mesons is the
same as exchanging the couple of quarks with the couple of
antiquarks. The factors arising from this exchange are the
$C$-parity operator eigenvalues~\cite{santopinto},
\begin{eqnarray}
C=(-1)^{L+J_{12}}.
\end{eqnarray}
The $G$-parity is a generalization of the concept of $C$-parity
such that members of an isospin multiplet can each be assigned a
good quantum number that would reproduce $C$-parity for the
neutral particle. The $G$-parity operator is defined as the
combination of $C$-parity and a $\pi$ rotation around the $y$ axis
in the isospin space~\cite{santopinto},
\begin{eqnarray}
G=C\mathcal{R}_y(\pi)=Ce^{i\pi I_2}.
\end{eqnarray}
The $G$-parity eigenstates are tetraquark states with flavor
charges equal to zero, i.e., strangeness equal to zero in the
light mesons case, and their eigenvalues are:
\begin{eqnarray}
G=(-1)^{L+J_{12}+I}.
\end{eqnarray}

\section{numerical results and discussions}

The diquark (antidiquark) is considered as a new compound object
$\bar{Q}$ ($Q$) with no internal spatial excitations, and spatial
excitations are assumed to occur only between $Q$ and $\bar{Q}$ in
the present numerical calculations, which results in that such a
tetraquark state has a lower energy than that of an internal
spatial excited ones. The orbital angular momentum $l_1$ and $l_2$
are therefore assumed to be zero. With these restrictions the
intermediate quantum number $J_{12}$ is the total spin angular
momentum $S$. The parity of a tetraquark with the
diquark-antidiquark structure is $P=(-1)^L$, the charge
conjugation is $C=(-1)^{L+S}$ and the $G$ parity is $G
=(-1)^{L+S+I}$.

Within the flux-tube model with the parameters fixed by fitting
the ordinary meson spectra~\cite{deng}, the convergent energies of
tetraquark states with this QCD quark cyclobutadiene and
diquark-antidiquark structures can be obtained by solving the
four-body Schr\"{o}dinger equation
\begin{eqnarray}
(H-E)\Phi^{q^2\bar{q}^2}_{IJ}=0,
\end{eqnarray}
with Rayleigh-Ritz variational principle by setting the numbers of
Gaussian wave functions $n_{1max}=n_{2max}=n_{3max}=6$. Minimum
and maximum ranges of the bases are 0.1 fm and 2.0 fm for Jacobi
coordinates $\mathbf{r}$, $\mathbf{R}$ and $\mathbf{X}$,
respectively. Quark contents and the corresponding masses with
specified quantum numbers, $I^GJ^{PC}$ or $IJ^P$, are shown in
Tables I-V, where $n$ stands for a non-strange quark ($u$ or $d$)
while $s$ stands for a strange quark, $E_I$ and $E_{II}$ represent
the energies of a QCD quark cyclobutadiene and a
diquark-antidiquark structure, respectively, the quantum number
$N$ denotes the total radial excitation. The $exotic$ in Table
I-II stands for a meson state which cannot be described by a
$q\bar{q}$ configuration.

The tetraquark states in the flux-tube model are generally lower
than that in the traditional quark models with additive two-body
confinement interaction with color factors used in early
multiquark state calculations~\cite{deng,scalar}. The reason is
that the multibody confinement potential can avoid the appearance
of the anti-confinement in a color symmetric quark or antiquark
pair. From Tables I-V, it can be seen that the two structures
generally give very close energies for tetraquark ground states.
However, the differences between two structures are about 40 MeV,
80 MeV and 120 MeV for spatial excitations between $Q$ and
$\bar{Q}$ with $L=1$, $L=2$ and $L=3$, respectively, which is
attributed to the reasons: (i) The expect values of the quadratic confinement
potentials linearly depend on the angular excitation
$L$~\cite{GEM}; (ii) The normal modes of the three 
quadratic confinement potentials of two flux-tube structures, 
see Eq.(7) and Eq.(8) are different. For a compact tetraquark state
in ground state, the separation among particles (quarks or
antiquarks) is generally smaller than one fm~\cite{ppbar}, so the
square of the length of each flux-tube is smaller than the length
itself. It is therefore predicted that the small difference of the
same quantum state between a linear confinement and a quadratic
one is about 50-80 MeV according to the calculations on hexaquark
states~\cite{ping}. Anyway, the differences of the ground states
between two structures are not big for both the linear and
quadratic confinement potentials.

In general, a tetraquark system should be the mixture of all
possible flux-tube structures. Such as in the process of a
meson-meson scattering, when two color singlet mesons are
separated far away, the dominant component of the system should be
two isolated color singlet mesons because other hidden color
flux-tube structures are suppressed due to the confinement. With
the separation reduction, a deuteron-like meson-meson molecular
state may be formed if the attractive force between two color
singlet mesons is strong enough. When they are close enough to be
within the range of confinement (about 1 fm), all possible
flux-tube structures including the QCD quark cyclobutadiene and
even more complicated flux-tube structures may appear due to the
excitation and rearrangements of flux-tubes and junctions.  All of
these hidden color components can not directly decay into two
colorful hadrons due to the color confinement. They must transform
back into two color singlet mesons by means of the rupture and
recombination of flux-tubes before decaying into two color singlet
mesons. The decay widths of these states are qualitatively
determined by the speed of the rupture and recombination of the
flux-tubes. This formation and decay mechanisms are similar to the
compound nucleus formation and therefore should induce a resonance
which is named as a ``color confined, multi-quark resonance"
state~\cite{resonance}. It is different from all of those
microscopic resonances discussed by S. Weinberg~\cite{weinberg}.
Bicudo and Cardoso studied tetraquark states using the triple
flip-flop potential including two meson-meson potentials and the
tetraquark four-body potential. They also found plausible the
existence of resonances in which the tetraquark component
originated by a flip-flop potential is the dominant
one~\cite{Bicudo}.

\begin{table}
\caption{The mass spectra for $nn\bar{n}\bar{n}$ states (unit:MeV).}
\begin{tabular}{cccccccccc}  \hline\hline
$I^GJ^{PC}$ & $N^{2S+1}L_J$  & $E_I$ && $E_{II}$&&States &PDG \\
\hline
$0^+0^{++}$ & $1^1S_0$ & 601  && 587  && $f_0(600)$  & 400--1200 \\
$0^+0^{++}$ & $2^1S_0$ & 1101 && 1019 && $f_0(980)$  & $980\pm 10$\\
$0^+1^{++}$ & $1^5D_1$ & 1927 && 1840 && $f_1(1285)$ & $1281.8\pm0.6$\\
$0^+1^{++}$ & $2^5D_1$ & 1984 && 1919 && $f_1(1420)$ & $1426.4\pm0.9$\\
$0^+1^{++}$ & $3^5D_1$ & 2373 && 2270 && $f_1(1510)$ & $1518\pm5$\\
$0^+2^{++}$ & $1^1D_2$ & 1328 && 1196 && $f_2(1270)$ & $1275.1\pm 1.2$\\
$0^+2^{++}$ & $2^1D_2$ & 1809 && 1614 && $f_2(1640)$ & $1639\pm 6$\\
$0^+2^{++}$ & $1^5S_2$ & 1468 && 1465 && $f_2(1430)$ & $\approx1430$\\
$0^+2^{++}$ & $2^5S_2$ & 1495 && 1508 && $f_2^{\prime}(1525)$ & $1525\pm5$\\
$0^+2^{++}$ & $1^5D_2$ & 1927 && 1840 && $f_2(1910)$ & $1903\pm 9$\\
$0^+2^{++}$ & $2^5D_2$ & 1984 && 1919 && $f_2(1950)$ & $1944\pm 12$\\
$0^+4^{++}$ & $2^5D_2$ & 1984 && 1919 && $f_4(2050)$ & $2018\pm 11$\\
$0^-2^{+-}$ & $1^3D_2$ & 1908 && 1836 && $exotic$    & ---\\
$0^+0^{-+}$ & $1^3P_0$ & 1624 && 1609 && $\eta(1295)$& $1294\pm4$\\
$0^+0^{-+}$ & $2^3P_0$ & 1656 && 1619 && $\eta(1405)$& $1409.8\pm2.5$\\
$0^+0^{-+}$ & $3^3P_0$ & 2063 && 2027 && $\eta(1475)$& $1476\pm4$\\
$0^+0^{-+}$ & $4^3P_0$ & 2097 && 2055 && $\eta(1760)$& $1756\pm9$\\
$0^+2^{-+}$ & $1^3P_2$ & 1624 && 1609 && $\eta_2(1645)$ & $1617\pm 5$\\
$0^-1^{--}$ & $1^1P_1$ & 1057 && 975  && $\phi(1020)$ & $1019.455\pm 0.020$\\
$0^-1^{--}$ & $2^1P_1$ & 1482 && 1358 && $\omega(1420)$& 1400--1450\\
$0^-1^{--}$ & $3^1P_1$ & 1583 && 1536 && $\omega(1650)$& $1670\pm30$\\
$0^-1^{--}$ & $1^5P_1$ & 1696 && 1651 && $\omega(1650)$& $1670\pm30$\\
$0^-3^{--}$ & $1^5P_3$ & 1696 && 1651 && $\omega_3(1670)$& $1667\pm 4$\\
$0^-1^{+-}$ & $1^3S_1$ & 1291 && 1304 && $h_1(1170)$   & $1170\pm20$\\
$0^-1^{+-}$ & $2^3S_1$ & 1391 && 1394 && $h_1(1380)$ & $1386\pm 19$\\
$1^-0^{++}$ & $1^1S_0$ & 1202 && 1210 && $a_0(980)$  & $980\pm20$\\
$1^-0^{++}$ & $2^3S_0$ & 1520 && 1528 && $a_0(1450)$ & $1474\pm 19$\\
$1^-1^{++}$ & $1^5D_1$ & 1927 && 1839 && $a_1(1260)$ & $1230\pm40$\\
$1^-1^{++}$ & $2^5D_1$ & 2373 && 2271 && $a_1(1640)$ &  $1647\pm22$\\
$1^-2^{++}$ & $1^5S_2$ & 1470 && 1467 && $a_2(1320)$ & $1318.3\pm0.6$\\
$1^-2^{++}$ & $1^1D_2$ & 1876 && 1807 && $a_2(1700)$ & $1732\pm16$\\
$1^+2^{+-}$ & $1^3D_2$ & 1910 && 1837 && $exotic$&---\\
$1^-0^{-+}$ & $1^3P_0$ & 1371 && 1307 && $\pi(1300)$   & $1300\pm 100$\\
$1^-1^{-+}$ & $1^3P_1$ & 1371 && 1307 && $\pi_1(1400)$ & $1354\pm 25$\\
$1^-1^{-+}$ & $1^5F_1$ & 1775 && 1691 && $\pi_1(1600)$ & $1662^{+15}_{-11}$\\
$1^+1^{--}$ & $1^1P_1$ & 1580 && 1558 && $\rho(1570)$& $1570\pm 36\pm 62$\\
$1^+1^{--}$ & $1^5F_1$ & 2157 && 2030 && $\rho(2150)$& $2149\pm17$\\
$1^+3^{--}$ & $1^5P_3$ & 1697 && 1651 && $\rho_3(1690)$& $1686\pm 4$\\
$1^+3^{--}$ & $2^5P_3$ & 2146 && 2062 && $\rho_3(1990)$& $1982\pm14$\\
$1^+1^{+-}$ & $1^3S_1$ & 1070 && 1089 && $b_1(1235)$&  $1229.5\pm3.2$\\
$2^+0^{++}$ & $1^1S_0$ & 1202 && 1211 && $exotic$& ---\\
$2^+0^{-+}$ & $1^3P_0$ & 1655 && 1617 && $exotic$& --- \\
$2^-1^{--}$ & $1^1P_1$ & 1580 && 1558 && $exotic$& --- \\
$2^-1^{--}$ & $1^5P_1$ & 1697 && 1651 && $exotic$& --- \\
$2^-1^{+-}$ & $1^3S_1$ & 1388 && 1391 && $exotic$& ---\\
$2^+1^{++}$ & $1^5D_1$ & 1927 && 1840 && $exotic$& --- \\
$2^+2^{++}$ & $1^1D_2$ & 1876 && 1807 && $exotic$& --- \\
$2^+2^{++}$ & $1^5S_2$ & 1468 && 1470 && $exotic$& --- \\
\hline \hline
\end{tabular}
\end{table}

Most tetraquark states in Tables I-V have the same quantum numbers
with ordinary meson states, and the calculated energies of many
tetraquark states are very close to the experimental data of the
mesons with the same quantum numbers~\cite{PDG}, especially states
with higher energy. This does not mean that the main component of
those experimental states must be tetraquark states. The fact is
that most of the experimentally observed mesons can be interpreted
as $q\bar{q}$ states (at least the main component) and
accommodated in the naive quark model, only a few of them may go
beyond $q\bar{q}$ configurations~\cite{salamanca,klempt}. However,
the calculations indicate that the tetraquark component in those
mesons, their energies are close to the tetraquark ones, can not
be excluded. This point is supported by the study on the nature of
scalar mesons~\cite{vijande}. Moreover the nucleon spin structure
study shows that even for the ground state the pentaquark
components $q^3q\bar{q}$ is indispensable in solving the proton
spin ``crisis"~\cite{dqing1,dqing2}. The strange magnetic momentum
of a nucleon originates from a strange sea quark $s\bar{s}$
component is nonzero~\cite{bszou}. So a comprehensive study of the
meson spectra must include the mixing of $q\bar{q}$ and
$qq\bar{q}\bar{q}$ Fock components and in turn requires the
knowledge of the off-shell interaction for annihilating or
creating a quark-antiquark pair into or from the vacuum. Quark
model should be unquenched and such an unquenched quark model
study is on going in our group.

With regard to nonstrange mesons, for some light $q\bar{q}$
excitation states, the orbital excitation energy between $q$ and
$\bar{q}$ may be higher than that of a quark-antiquark pair
excited from the quark sea, so these meson states prefer to have
high Fock component $qq\bar{q}\bar{q}$. Such as the meson
$\sigma$, it can be described as the ground state with the quark
content $nn\bar{n}\bar{n}$ rather than the excited states of a
$q\bar{q}$ meson~\cite{deng}, which is consistent with many other
works~\cite{alford,maiani2,Pelaez1,Pelaez2}. The first radial
excited state of the $nn\bar{n}\bar{n}$ state is very close to the
experimental value of the meson $f_0(980)$, the tetraquark state
$nn\bar{n}\bar{n}$ may therefore be one of the main components,
which is supported by Vijande's work on the nature of scalar
mesons~\cite{vijande}. The decay of the meson $f_0(980)$ into
$K\bar{K}$ can be accounted for by other strangeness components,
such as $s\bar{s}$ and $ns\bar{n}\bar{s}$. The meson $f_0(1500)$
can not be described as a $q\bar{q}$ meson, the mass and decay are
compatible with it being the ground state glueball mixed with the
nearby states of the $0^{++}$ $\bar{q}q$ nonet~\cite{klempt}. In
the quark models, another interpretation of the meson $f_0(1500)$
is that the main component might be a tetraquark state
$ns\bar{n}\bar{s}$~\cite{vijande}. The meson $f_2(1430)$ has no
proper member in the $q\bar{q}$ picture either~\cite{salamanca}.
It is suggested that the main component is a tetraquark
$nn\bar{n}\bar{n}$ with quantum numbers $1^5S_2$ in the flux-tube
model. This state is not confirmed in the PDG and even recent
measurements have suggested a different assignment of quantum
numbers, which could make it compatible with the lightest scalar
glueball~\cite{morningstar}.

With respect to $I=\frac{1}{2}$ strange mesons, most of them can
be interpreted as the states which dominated by $q\bar{q}$ components
in the quark models except for three mesons $\kappa(800)$, $K^*(1410)$ and
$K_2(1580)$~\cite{salamanca}. For the same reason with the meson
$\sigma$, our model recommends a ground tetraquak state
$nn\bar{n}\bar{s}$ $1^1S_0$ with energy close to the meson
$\kappa(800)$, which is compatible with other
works~\cite{alford,maiani2,Pelaez1,Pelaez2}. For the meson
$K^*(1410)$, its assignment to the $2^3S_1$ state of the meson
$K^*(892)$ is not only excluded by the large mass difference, but
also by its decay modes~\cite{salamanca}. A possible
interpretation of the main component of this state is a tetraquark
state $nn\bar{n}\bar{s}$ with quantum numbers $1^1P_1$ instead of
a pure $q\bar{q}$ pair. The meson $K_2(1580)$ has also no proper
member in the $q\bar{q}$ spectra~\cite{salamanca}, our tetraquark
state $1^3P_2$ mass is a little lower than experimental data. This
state is clearly uncertain, it was reported in only one
experimental work more than twenty years ago and has never been
measured again.

Concerning the exotic meson sector, the quantum numbers rule out
the pure $q\bar{q}$ possibility. The $\pi_1$ mesons of
$I^GJ^{PC}=1^-1^{-+}$are listed as manifestly exotic states by
several experiments~\cite{lu,yao,admas,nozar}. Many theoretical
studies have been made and various interpretations were proposed:
hybrid meson states~\cite{page,chetyrkin,jin,bernard}, $\pi\eta$
molecular states~\cite{zhangr} and tetraquark
states~\cite{hxchen1,hxchen2}. Two mesons $\pi_1(1400)$ and
$\pi_1(1600)$ are studied in the flux-tube model, see Table I,
which indicates that the main components of $\pi_1(1400)$ and
$\pi_1(1600)$ might be tetraquark state $nn\bar{n}\bar{n}$ with
quantum numbers $1^3P_1$ and $1^5F_1$, respectively. The
tetraquark states $ns\bar{n}\bar{s}$ with quantum numbers $I=0,1$
and $J^{PC}=1^{-+}$ are predicted in the flux-tube model, the
energies are around 1850 MeV (see Table II) which is consistent
with the predictions on $J^{PC}=1^{-+}$ tetraquark states in the
QCD sum rule~\cite{hxchen1,hxchen2}. The exotic meson states with
quantum numbers $J^{PC}=2^{+-}$ are predicted in the tetraquark
picture, the states $nn\bar{n}\bar{n}$ and $ns\bar{n}\bar{s}$ have
the lowest masses around 1880 MeV and 2100 MeV, respectively. In
addition, many flavor $I=2$ exotic meson states are also
calculated for further studies, see Table I.

\begin{table}
\caption{The mass spectra for $ns\bar{n}\bar{s}$ states (unit:MeV).}
\begin{tabular}{ccccccccc}\hline\hline
$I^GJ^{PC}$ & $N^{2S+1}L_J$ & $E_I$  && $E_{II}$ && States &PDG\\
\hline
$0^+0^{++}$ & $1^1S_0$ & 1316 && 1318 && $f_0(1370)$& 1200--1500\\
$0^+0^{++}$ & $2^1S_0$ & 1583 && 1590 && $f_0(1500)$& $1505\pm 6$\\
$0^+0^{++}$ & $3^1S_0$ & 1676 && 1661 && $f_0(1710)$& $1720\pm 6$\\
$0^+0^{++}$ & $1^5D_0$ & 2174 && 2095 && $f_0(2100)$& $2103\pm 8$\\
$0^+0^{++}$ & $1^5D_0$ & 2174 && 2095 && $f_0(2200)$& $2189\pm 13$\\
$0^+2^{++}$ & $1^5S_2$ & 1751 && 1755 && $f_2(1810)$& $1815\pm 12$\\
$0^+2^{++}$ & $1^1D_2$ & 2033 && 1946 && $f_2(2010)$& $2011^{+62}_{-76}$\\
$0^+2^{++}$ & $2^1D_2$ & 2141 && 2073 && $f_2(2150)$& $2157\pm 12$\\
$0^+2^{++}$ & $1^5D_2$ & 2174 && 2095 && $f_2(2150)$& $2157\pm 12$\\
$0^+0^{-+}$ & $1^3P_0$ & 1867 && 1831 && $\eta(1760)$& $1756\pm9$\\
$0^+2^{-+}$ & $1^3P_2$ & 1867 && 1831 && $\eta_2(1870)$& $1842\pm 8$\\
$0^-1^{--}$ & $1^1P_1$ & 1773 && 1740 && $\phi(1680)$& $1680\pm 20$\\
$0^-1^{--}$ & $2^1P_1$ & 1892 && 1866 && ---& ---\\
$0^-1^{+-}$ & $1^3S_1$ & 1583 && 1586 && $h_1(1595)$& $1594\pm 15^{+10}_{-60}$\\
$0^-1^{+-}$ & $2^3S_1$ & 1626 && 1628 && ---& ---\\
$0^+1^{-+}$ & $1^3P_1$ & 1865 && 1828 && $exotic$& ---\\
$0^-2^{+-}$ & $1^3D_2$ & 2108 && 2076 && $exotic$& ---\\
$0^-3^{--}$ & $1^5P_3$ & 1968 && 1928 && $\phi_3(1850)$&$1854\pm7$\\
$1^-0^{++}$ & $1^1S_0$ & 1320 && 1318 && $a_0(980)$& $980\pm20$\\
$1^-0^{++}$ & $2^1S_0$ & 1584 && 1590 && $a_0(1450)$& $1474\pm19$\\
$1^-2^{++}$ & $1^5S_2$ & 1751 && 1755 && $a_2(1700)$& $1732\pm 16$\\
$1^-2^{++}$ & $1^1D_2$ & 2033 && 1945 && ---& ---\\
$1^-0^{-+}$ & $1^3P_0$ & 1867 && 1831 && $\pi(1800)$& $1816\pm 14$\\
$1^-1^{-+}$ & $1^3P_1$ & 1867 && 1831 && $exotic$& ---\\
$1^-2^{+-}$ & $1^3D_2$ & 2108 && 2076 && $exotic$& ---\\
$1^-2^{-+}$ & $1^3P_2$ & 1867 && 1831 && $\pi_2(1880)$& $1895\pm 16$\\
$1^-2^{-+}$ & $1^3F_2$ & 2309 && 2186 && $\pi_2(2100)$& $2090\pm 29$\\
$1^+1^{--}$ & $1^1P_1$ & 1772 && 1739 && $\rho(1700)$& $1700\pm 20$\\
$1^+1^{--}$ & $2^1P_1$ & 1892 && 1866 && $\rho(1900)$& $1909\pm17\pm25$\\
$1^+1^{--}$ & $1^5F_5$ & 2376 && 2259 && $\rho(2150)$& $2149\pm17$\\
$1^+3^{--}$ & $1^5P_3$ & 1967 && 1928 && $\rho_3(1990)$& $1982\pm 14$\\
$1^+3^{--}$ & $1^1F_3$ & 2248 && 2117 && $\rho_3(2250)$& $\sim2232$\\
$1^+5^{--}$ & $1^5F_5$ & 2376 && 2259 && $\rho_5(2350)$& $2330\pm 35$\\
$1^-1^{++}$ & $1^5D_1$ & 2173 && 2095 && ---& ---\\
$1^+1^{+-}$ & $1^3S_1$ & 1583 && 1586 && ---& ---\\
\hline\hline
\end{tabular}
\end{table}

\begin{table}
\caption{The mass spectra for $ss\bar{s}\bar{s}$ states.}
\begin{tabular}{cccccccc} \hline\hline
$I^GJ^{PC}$ & $N^{2S+1}L_J$ & $E_I$ && $E_{II}$ && States&PDG\\
\hline
$0^+0^{++}$ & $1^1S_0$ & 1919 && 1925 && $f_0(2020)$&$1992\pm16$\\
$0^+0^{++}$ & $1^5D_0$ & 2440 && 2365 && $f_0(2330)$&$2314\pm25$\\
$0^+2^{++}$ & $1^5S_2$ & 2051 && 2044 && $f_2(2010)$&$2011^{+62}_{-76}$\\
$0^+2^{++}$ & $1^1D_2$ & 2423 && 2354 && $f_2(2300)$&$2297\pm28$\\
$0^+2^{++}$ & $1^5D_2$ & 2440 && 2365 && $f_2(2300)$&$2297\pm28$\\
$0^+2^{++}$ & $1^1D_2$ & 2423 && 2354 && $f_2(2340)$&$2340\pm55$\\
$0^+2^{++}$ & $1^5D_2$ & 2440 && 2365 && $f_2(2340)$&$2340\pm55$\\
$0^+4^{++}$ & $1^5D_2$ & 2440 && 2365 && $f_4(2300)$&$\sim2314$\\
$0^-1^{--}$ & $1^1P_1$ & 2201 && 2176 && $\phi(2170)$&$2175\pm15$\\
$0^+0^{-+}$ & $1^3P_0$ & 2232 && 2195 && $\eta(2225)$&$2226\pm16$\\
$0^-1^{--}$ & $1^5P_1$ & 2249 && 2209 && $\phi(2170)$&$2175\pm15$\\
$0^-1^{+-}$ & $1^3D_1$ & 2432 && 2359 && ---& ---\\
\hline\hline
\end{tabular}
\end{table}

\begin{table}
\caption{The mass spectra for $nn\bar{n}\bar{s}$ states.}
\begin{tabular}{cccccccc} \hline\hline
$IJ^{P}$ & $N^{2S+1}L_J$ & $E_I$ && $E_{II}$  && States &PDG\\
\hline
$\frac{1}{2}0^{+}$ & $1^1S_0$ & 995  && 947  && $K^*_0(800)$&$676\pm40$\\
$\frac{1}{2}0^{+}$ & $2^1S_0$ & 1383 && 1380 && $K^*_0(1430)$& $1425.6\pm 1.5$\\
$\frac{1}{2}0^{+}$ & $1^5D_0$ & 2050 && 1968 && $K^*_0(1950)$& $1945\pm 10\pm 20$\\
$\frac{1}{2}2^{+}$ & $1^5D_2$ & 2050 && 1968 && $K^*_2(1980)$& $1973\pm 8\pm 25$\\
$\frac{1}{2}4^{+}$ & $1^5D_4$ & 2050 && 1968 && $K^*_4(2045)$& $2045\pm 9$\\
$\frac{1}{2}0^{-}$ & $1^3P_0$ & 1514 && 1451 && $K(1460)$&$\sim1460$\\
$\frac{1}{2}0^{-}$ & $2^3P_0$ & 1739 && 1697 && $K(1630)$& $1629\pm 7$\\
$\frac{1}{2}0^{-}$ & $3^3P_0$ & 1772 && 1754 && $K(1830)$& $\sim 1830$\\
$\frac{1}{2}1^{-}$ & $1^1P_1$ & 1430 && 1367 && $K^*(1410)$& $1414\pm 15$\\
$\frac{1}{2}1^{-}$ & $2^1P_1$ & 1709 && 1666 && $K^*(1680)$& $1717\pm 27$\\
$\frac{1}{2}1^{+}$ & $1^3S_1$ & 1254 && 1233 && $K_1(1270)$& $1272\pm 7$\\
$\frac{1}{2}1^{+}$ & $2^3S_1$ & 1447 && 1456 && $K_1(1400)$& $1403\pm 7$\\
$\frac{1}{2}1^{+}$ & $1^3D_1$ & 1749 && 1644 && $K_1(1650)$& $1650\pm 50$\\
$\frac{1}{2}2^{+}$ & $1^5S_2$ & 1603 && 1601 && $K_2^*(1430)$& $1425.6\pm1.5$\\
$\frac{1}{2}2^{+}$ & $1^1D_2$ & 1685 && 1573 && $K_2^*(1430)$& $1425.6\pm1.5$\\
$\frac{1}{2}2^{+}$ & $2^1D_2$ & 2014 && 1942 && $K_2^*(1980)$& $1973\pm8\pm25$\\
$\frac{1}{2}2^{-}$ & $1^3P_2$ & 1514 && 1451 && $K_2(1580)$& $\sim 1580$\\
$\frac{1}{2}2^{-}$ & $1^5P_2$ & 1828 && 1786 && $K_2(1770)$& $1773\pm 8$\\
$\frac{1}{2}2^{-}$ & $1^5P_2$ & 1828 && 1786 && $K_2(1820)$& $1816\pm 13$\\
$\frac{1}{2}3^{-}$ & $1^5P_3$ & 1828 && 1786 && $K_3^*(1780)$& $1776\pm 7$\\
\hline\hline
\end{tabular}
\end{table}

\begin{table}
\caption{The mass spectra for $ns\bar{s}\bar{s}$ states.}
\begin{tabular}{cccccccc} \hline\hline
$IJ^{P}$ & $N^{2S+1}L_J$ & $E_I$ && $E_{II}$ && States&PDG\\
\hline
$\frac{1}{2}0^{+}$ & $1^1S_0$ & 1757 &&  1762 &&---& ---\\
$\frac{1}{2}0^{+}$ & $2^1S_0$ & 1938 &&  1945 &&$K^*_0(1950)$& $1945\pm10\pm20$\\
$\frac{1}{2}3^{+}$ & $1^5D_3$ & 2308 &&  2230 &&$K_3(2320)$& $2324\pm 24$\\
$\frac{1}{2}0^{-}$ & $1^3P_0$ & 2026 &&  1984 &&---& ---\\
$\frac{1}{2}0^{-}$ & $2^3P_0$ & 2088 &&  2051 &&---& ---\\
$\frac{1}{2}1^{-}$ & $1^1P_1$ & 2051 &&  2024 &&---& ---\\
$\frac{1}{2}1^{-}$ & $2^1P_1$ & 2160 &&  2139 &&---& ---\\
$\frac{1}{2}2^{-}$ & $1^5P_2$ & 2108 &&  2068 &&$K_2^*(2250)$& $2247\pm17$\\
$\frac{1}{2}4^{-}$ & $1^5F_4$ & 2503 &&  2386 &&$K_4^*(2500)$& $2490\pm20$\\
$\frac{1}{2}5^{-}$ & $1^5F_5$ & 2503 &&  2386 &&$K_5^*(2380)$& $2382\pm14\pm19$\\
$\frac{1}{2}1^{+}$ & $1^3S_1$ & 1778 &&  1774 &&---& ---\\
$\frac{1}{2}1^{+}$ & $2^3S_1$ & 1864 &&  1862 &&---& ---\\
$\frac{1}{2}2^{+}$ & $1^5S_2$ & 1904 &&  1900 &&---& ---\\
$\frac{1}{2}2^{+}$ & $1^1D_2$ & 2284 &&  2215 &&---& ---\\
$\frac{1}{2}2^{+}$ & $2^1D_2$ & 2440 &&  2386 &&---& ---\\
\hline\hline
\end{tabular}
\end{table}

\section{summary}

The QCD quark cyclobutadiene, a new flux-tube structure, is
proposed in the framework of the flux-tube model. The flux-tube
ring in the QCD quark cyclobutadiene can be described as a
glueball, four quarks are connected to the flux-tube ring by four
fundamental flux-tubes, thus the QCD quark cyclobutadiene can be
viewed as a $qq\bar{q}\bar{q}$-glueball hybrid. It provides a new
sample for understanding the structures of exotic hadrons. The
three familiar flux-tube structures can be taken as the ground
states of a tetraquark system, the QCD quark cyclobutadiene may be
an excited state which is obtained by means of creating Y-shaped
junctions and flux-tubes from the vacuum and the rearrangement of
some flux-tubes. The QCD quark cyclobutadiene and other three
flux-tube structures are QCD isomeric compounds due to the same
quark component and different flux-tube structures.

Most meson states in the PDG can be described as $q\bar{q}$
configurations, only a few meson states, $\sigma$, $\kappa(800)$,
$f_0(980)$, $f_0(1500)$, $\pi_1(1400)$, $\pi_1(1600)$, $f_2(1430)$
and $K^*(1410)$, are difficult to be interpreted as $q\bar{q}$
mesons. The tetraquark state as their main components is one of
possible interpretations of their flavor components. Some exotic
meson states as tetraquark states are predicted in the flux-tube
model. The tetraquark states $ns\bar{n}\bar{s}$ with quantum
numbers $I=0,1$ and $J^{PC}=1^{-+}$ have the lowest masses around
1850 MeV. The tetraquark states $nn\bar{n}\bar{n}$ and
$ns\bar{n}\bar{s}$ with quantum numbers $J^{PC}=2^{+-}$ have the
lowest masses around 1880 MeV and 2100 MeV, respectively. The
tetraquark states with $I=2$ are also predicted in the flux-tube
model.

Even though up to now no tetraquark state has been well
established experimentally. It is indispensable to continue the
study of the tetraquark system because the tetraquark component in
mesons can not be ruled out and may play an important role in the
properties of mesons, similar to the fact that the pentaquark
components play an important role even in the nucleon ground
state.

The tetraquark states, if they really exist, should be the
mixtures of all kinds of flux-tube structures which can transform
one another. In this way, the flip-flop of flux-tube structures
can induce a resonance which is named as a ``color confined,
multi-quark resonance" state. To verify such a new resonance is
not easy. We admit that this analysis is based on the mass
calculation only, the crucial test of the components of exotic
mesons is determined by the systematic study of their decays,
which involves channel coupling calculation containing all
possible flux-tube structures and mixing between $q\bar{q}$ and
tetraquark components and so much more information of low energy
QCD are needed. 

\acknowledgments{This work is supported partly by the National
Science Foundation of China under Grant Nos. 11047140, 11035006,
11175088 and the PhD Program Funds of Chongqing Jiaotong
University.}


\begin{references}
\bibitem{godfrey} S. Godfrey and J. Napolitano, Rev. Mod. Phys. \textbf{71}, 1411 (1999).
\bibitem{amsler} C. Amsler, N. A. T\"{o}rnqvist, Phys. Re. \textbf{389}, 61 (2004).
\bibitem{jaffe} R.L. Jaffe, Phys. Rep. \textbf{409}, 1 (2005).
\bibitem{aubert} B. Aubert et al. (BABAR Collaboration), Phys. Rev. Lett. \textbf{90}, 242001 (2003).
\bibitem{choi} S.K. Choi et al. (Belle Collaboration), Phys. Rev. Lett., 262001 (2003)
\bibitem{evdokimov} A.V. Evdokimov et al. (SELEX Collaboration), Phys. Rev. Lett. \textbf{93}, 242001 (2004).
\bibitem{thompson} D.R. Thompson et al. (E852 Collaboration), Phys. Rev. Lett. \textbf{79}, 1630 (1997).
\bibitem{aele} A. Abele et al. (Crystal Barrel Collaboration), Phys. Lett. B \textbf{446}, 349 (1999).
\bibitem{adams} G.S. Adams et al. (E862 Collaboration), Phys. Lett. B \textbf{657}, 27 (2007).
\bibitem{alekseev} M.G. Alekseev, {em et al.} (COMPASS Collaboration), Phys. Rev. Lett. \textbf{104}, 241803 (2010).
\bibitem{nerling} F. Nerling, {em et al.} (COMPASS Collaboration), Arxiv: 1208.0474v1 (hep-ex).
\bibitem{wong} C.Y. Wong, Phys. Rev. D \textbf{69},055202 (2004).
\bibitem{close} F.E. Close and P.R. Page, Phys. Lett. B \textbf{578}, 119 (2004).
\bibitem{swanson} E.S. Swanson, Phys. Lett. B \textbf{588}, 189 (2004).
\bibitem{tornqvist} N.A. Tornqvist, Phys. Lett. B \textbf{590}, 209 (2004).
\bibitem{maiani1} L. Maiani, F. Piccinini and A.D. Polosa, et. al., Phys. Rev. D \textbf{71}, 014028 (2005).
\bibitem{hogaasen} H. Hogaasen, J.M. Richard and P. Sorba, Phys. Rev. D \textbf{73}, 054013 (2006).
\bibitem{ebert1} D. Ebert, R.N. Faustov and V.O. Galkin, Phys. Lett. B \textbf{634}, 214 (2006).
\bibitem{barnea} N. Barnea, J. Vijande and A. Valcarce, Phys. Rev. D \textbf{73}, 054004 (2006).
\bibitem{vijande1} J. Vijande, E. Weissman and N. Barnea, et al., Phys. Rev. D\textbf{76}, 094022 (2007).
\bibitem{janc} D. Janc and M. Rosina, Few-Body Systems \textbf{35}, 175-196 (2004).
\bibitem{olson} S.L. Olson, Nucl. Phys. A \textbf{827}, 53c (2009), and references therein.
\bibitem{hxchen} H.X. Chen, A. Hosaka, and S.L. Zhu, Phys. Rev. D \textbf{78}, 054017 (2008).
\bibitem{ckjiao} C.K. Jiao, W. Chen, H.X. Chen and S.L. Zhu, Phys. Rev. D \textbf{79}, 114034 (2009).
\bibitem{dudek}  J.J. Dudek, R.G. Edwards, M.J. Peardon, D.G. Richards and C.E. Thomas, Phys. Rev. Lett. \textbf{103}, 262001 (2009).
\bibitem{dmitrasinovic} V. Dmitrasinovic, Phys. Rev. D \textbf{67}, 114007 (2003).
\bibitem{alexandrou} C. Alexandrou, P. De Forcrand, and A. Tsapalis, Phys. Rev. D \textbf{65}, 054503 (2002).
\bibitem{takahashi} T.T. Takahashi, H. Suganuma, Y. Nemoto and H. Matsufuru, Phys. Rev. D \textbf{65}, 114509 (2002).
\bibitem{okiharu1} F. Okiharu, H. Suganuma and T.T. Takahashi, Phys. Rev. D\textbf{72}, 014505 (2005).
\bibitem{okiharu2} F. Okiharu, H. Suganuma and T.T. Takahashi, Phys. Rev. Lett.\textbf{94}, 192001 (2005).
\bibitem{anderson} P.W. Anderson, Phys. Today, \textbf{53}, No.2, 11 (2000);
\bibitem{teng} F. Wang, G.H. Wu, L.J. Teng and T. Goldman, Phys. Rev. Lett.\textbf{69}, 2901 (1992).
\bibitem{barnes}  T. Barnes, F.E. Close and H.J. Lipkin, Phys. Rev. D \textbf{68}, 054006 (2003).
\bibitem{lee} I.W. Lee, A. Faessler, T. Gutsche and V.E. Lyubovitskij, Phys.Rev. D \textbf{80}, 094005 (2009).
\bibitem{xliu} X. Liu, X.Q. Zeng and X.Q. Li, Phys. Rev. D \textbf{72}, 054023(2005).
\bibitem{jaffe1} R.L. Jaffe and F. Wilczek, Phys. Rev. Lett. \textbf{91}, 232003 (2003).
\bibitem{maiani2} L. Maiani, F. Piccinini, A.D. Polosa and V. Riquerx, Phys. Rev. Lett. \textbf{93}, 212002 (2004).
\bibitem{ebert2} D. Ebert, R.N. Faustov, V.O. Galkin and W. Lucha, Phys. Rev. D \textbf{76}, 114015 (2007).
\bibitem{hxhuang} H.X. Huang, C.R. Deng and J.L. Ping et al, Phys. Rev. C \textbf{77}, 025201 (2008).
\bibitem{jlping} J.L. Ping, H.X. Huang and C.R. Deng, et al, Phys. Rev. C \textbf{79}, 065203 (2009).
\bibitem{ping}  J.L. Ping, C.R. Deng, F. Wang and T. Goldman, Phys. Lett. B \textbf{659}, 607(2008).
\bibitem{ppbar} C.R. Deng,  J.L. Ping, Y.C. Yang and F. Wang, Phys. Rev. D \textbf{86}, 014008 (2012).
\bibitem{deng} C.R. Deng, J.L. Ping, F. Wang and T. Goldman, Phys. Rev. D \textbf{82}, 074001 (2010).
\bibitem{isgur} N. Isgur and Jack. Paton, Phys. Rev. D \textbf{31}, 2190 (1985).
\bibitem{Bali} G.S. Bali, Phys. Rev. D \textbf{62}, 114503 (2000).
\bibitem{koma} Y. Koma, H. Suganuma and H. Toki, Phys. Rev. D \textbf{60}, 074024 (1999).
\bibitem{maris} P. Maris and C.R. Roberts, Int. J. Mod. Phys. E \textbf{12}, 297 (2003).
\bibitem{ishii} N. Ishii, S. Aoki, and T. Hatsuda, Phys. Rev. Lett. \textbf{99}, 022001 (2007).
\bibitem{takahashi1} T.T. Takahashi and Y. Kanada-En$^,$yo, Phys. Rev. D \textbf{82}, 094506 (2010).
\bibitem{inoue} T. Inoue, N. Ishii and S. Aoki et al (HAL QCD Collaboration), Phys. Rev. Lett. \textbf{106}, 162002 (2011).
\bibitem{ishii1} N. Ishii, AIP Conf. Proc. \textbf{1355}, 206-213 (2011).
\bibitem{feinberg} G. Feinberg and J. Sucher, Phys. Rev. D \textbf{20}, 1717 (1979).
\bibitem{greenberg} O.W. Greenberg and H.J. Lipkin, Nucl. Phys. A \textbf{370}, 349 (1981).
\bibitem{weinstein1} J. Weinstein and N. Isgur, Phys. Rev. Lett. \textbf{48}, 659 (1982).
\bibitem{weinstein} J. Weinstein and N. Isgur, Phys. Rev. D \textbf{41}, 2238 (1990).
\bibitem{oka1} M. Oka, Phys. Rev. D \textbf{31}, 2274 (1985).
\bibitem{oka2} M. Oka and C.J. Horowitz, Phys. Rev. D \textbf{31}, 2773 (1985).
\bibitem{karliner} M. Karliner and H.J. Lipkin, Phys. Lett. B \textbf{575}, 249 (2003).
\bibitem{vijande2} J. Vijande, A. Valcarce, and J.-M. Richard, Phys. Rev. D \textbf{76}, 114013 (2007).
\bibitem{ww} F. Wang and C.W. Wong, Nuovo Cimento A \textbf{86}, 283 (1985).
\bibitem{goldman} T. Goldman and S. Yankielowicz, Phys. Rev. D \textbf{12}, 2910 (1975).
\bibitem{Nawa} M. Iwasaki, S. Nawa, T. Sanada and F. Takagi, Phys. Rev. D \textbf{68}, 074007 (2003).
\bibitem{carlson} J. Carlson and V.R. Pandharipande, Phys. Rev. D \textbf{43}, 1652 (1991).
\bibitem{GEM} E. Hiyama, Y. Kino, and M. Kamimura, Prog. Part. Nucl. Phys. \textbf{51} 223 (2003).
\bibitem{Llanes-Estrada} F.J. Llanes-Estrada, ECONF c0309101, FRWP011 (2003); ArXiv: 0311235 (hep-ph).
\bibitem{santopinto} E. Santopinto and Giuseppe Galat$\grave{a}$, Phys. Rev. C \textbf{75}, 045206 (2007).
\bibitem{scalar} C.R. Deng, J.L. Ping, and F. wang, ArXiv: 1204.0246 (hep-ph).
\bibitem{resonance} F. Wang, J.L. Ping, H.R. Pang and L.Z. Chen, Nucl. Phys. A \textbf{790}, 493c (2007).
\bibitem{weinberg} S. Weinberg, {\it The Quantum Theory of Fields}, (Combridge University Press, 1995), V.I, p.159.
\bibitem{Bicudo} P. Bicudo and M. Cardoso, Phys. Rev. D \textbf{83}, 094010 (2011).
\bibitem{PDG} K. Nakamura {\em et al}, [Particle Data Group], J. Phys. G \textbf{37}, 075021 (2010).
\bibitem{salamanca} J. Vijande, F. Fernandez and A. Valcarce, J. Phys. G \textbf{31}, 481 (2005).
\bibitem{klempt} E. Klempt and A. Zaitsev, Phys. Rep. \textbf{454}, 1 (2007).
\bibitem{vijande} J. Vijande, A. Valcarce, F. Fernandez and B. Silvestre-Brac, Phys. Rev. D \textbf{72} 034025 (2005).
\bibitem{dqing1} D. Qing, X.S. Chen, and F. Wang, Phys. Rev. C \textbf{57}, 31 (1998).
\bibitem{dqing2} D. Qing, X.S. Chen, and F. Wang, Phys. Rev. D \textbf{58}, 114032 (1998).
\bibitem{bszou} B.S. Zou and D.O. Riska, Phys. Rev. Lett. \textbf{95}, 072001 (2005).
\bibitem{alford} M.G. Alford and R.L. Jaffe, Nucl. Phys. B \textbf{578}, 367 (2000).
\bibitem{Pelaez1} J.R. Pelaez, Phys. Rev. Lett. \textbf{92}, 102001 (2004).
\bibitem{Pelaez2} J.R. Pelaez and G. Rios, Phys. Rev. Lett. \textbf{97}, 242002 (2006).
\bibitem{morningstar} C.J. Morningstar and M. Peardon, Phys. Rev. D \textbf{60} 034509 (1999).
\bibitem{lu} M. Lu et al. (E852 Collaboration), Phys. Rev. Lett. \textbf{94}, 032002 (2005).
\bibitem{yao} W.M. Yao et al. (Particle Data Group), J. Phys. G \textbf{33}, 1 (2006).
\bibitem{admas} G.S. Adams et al. (E862 Collaboration), Phys. Lett. B \textbf{657}, 27 (2007).
\bibitem{nozar} M. Nozar et al. (CLAS Collaboration), Phys. Rev. Lett. \textbf{102}, 102002 (2008).
\bibitem{page} P.R. Page, E.S. Swanson, and A.P. Szczepaniak, Phys. Rev. D \textbf{59}, 034016 (1999).
\bibitem{chetyrkin} K.G. Chetyrkin and S. Narison, Phys. Lett. B \textbf{485}, 145 (2000).
\bibitem{jin} H.Y. Jin, J.G. Korner, and T.G. Steele, Phys. Rev. D \textbf{67}, 014025 (2003).
\bibitem{bernard} C. Bernard et al., Phys. Rev. D \textbf{68}, 074505 (2003).
\bibitem{zhangr} R. Zhang, Y.B. Ding, X.Q. Li, and P.R. Page, Phys. Rev. D \textbf{65}, 096005 (2002).
\bibitem{hxchen1} H.X. Chen, A. Hosaka, and S.L. Zhu, Phys. Rev. D \textbf{78}, 054017 (2008).
\bibitem{hxchen2} H.X. Chen, A. Hosaka, and S.L. Zhu, Phys. Rev. D \textbf{78}, 117502 (2008).
\end{references}
\end{document}